\documentclass[11pt]{article}
\usepackage[paper=a4paper,dvips,top=1.5cm,left=1.5cm,right=1.5cm,
    foot=1cm,bottom=1.5cm]{geometry}
\usepackage{times}
\usepackage{graphicx}
\usepackage{amsmath}
\usepackage{amsfonts}
\date{}

\title{Violation of Bell Inequalities as a Violation of Fair Sampling\\in Threshold Detectors}

\author{Guillaume Adenier\\
    \small
        \emph{International Center for Mathematical Modeling in Physics
Engineering, Economy and Cognitive Sc.}\\\small
        \emph{University of V\"{a}xj\"{o}, Vejdes Plats 6, 35195 V\"{a}xj\"{o},
Sweden}}
\begin{document}
% Start your text
\maketitle

\begin{abstract}
Photomultiplier tubes and avalanche photodiodes, which are
commonly used in quantum optic experiments, are sometimes referred
to as threshold detectors because, in photon counting mode, they
cannot discriminate the number of photoelectrons initially
extracted from the absorber in the detector. We argue that they
can be called threshold detectors on more account than that. We
point out that their their functioning principle relies on two
thresholds that are usually thought unimportant individually in
the context of EPR-Bell discussion. We show how the combined
effect of these threshold can lead to a significant sampling
selection bias in the detection of pairs of pulses, resulting in
an apparent violation of Bell inequalities.
\end{abstract}

\section{Introduction}
As much as EPR-Bell experiments have consistently shown results in
agreement with the predictions of quantum mechanics, they have
however never been able to entirely remove the possibility of
local realistic models explaining the observed correlations.
Various loopholes that render the probability space explicitly
contextual have been identified, and it is the concern of
experimenters to make sure that these loopholes cannot be
responsible for the observed violation of Bell inequalities. The
derivation of a Bell inequality \cite{Bell64} requires indeed that
an experiment involving several incompatible measurements can be
written on a single probability space, independently of the
measurement context. This has been criticized by a number of
authors \cite{Accardi81,Accardi2,Accardi3,Hess1,Hess2}, and is at
the heart of the V\"axj\"o Interpretation of Quantum Mechanics
\cite{KHR1a,KHR1b,KHR1c,KHR2,KHR3,KHR4a,KHR4b,KHR10,KHR05}---which
is a contextual statistical realistic interpretation of Quantum
Mechanics.

The main class of contextualizing loopholes that have been
thoroughly addressed in EPR experiments are the ones based on a
communication between the parties. By rapidly changing the
measurement settings during the flight of the particles,
experimenters have excluded the possibility of an exchange of
information between the parties through a classical channel. The
first experiment to thus enforce a space-like interval between the
remote measurements was that of Aspect \emph{et al}
\cite{Aspect82} by means of a periodical and asynchronous
switching between the measurement settings. The concept was
furthered by Weihs \emph{et al} \cite{Weihs} with a random fast
switching, and it was extended recently to separating the
measurement settings from the emission at the source by a
space-like interval \cite{Scheidl}.

The other main class of contextualizing loopholes arises when an
experiment is performed with a low overall efficiency\footnote{An
experiment with ion traps \cite{Rowe} did reach perfect detection
efficiency. It could not however fulfill the locality condition
and was thus powerless to challenge local realism the way EPR-Bell
experiments are meant to.}. The correlations are therefore not
measured on the population of all emitted pairs, as a derivation
of a Bell inequality would have it, but on a small sample instead.
The trouble occurs if the sampling process itself is unfair, that
is, if the probability for a given particle to be detected in
either channel depends on the local measurement setting. Then the
subset $\Omega_\alpha$ (resp. $\Omega_\beta$) of the sample space
$\Omega$ that is spanned locally by the single events detected by
Alice (resp. by Bob) depends explicitly on the local measurement
context $\alpha$ (resp. $\beta$). At first sight, such a
contextuality might appear harmless since it is fully local.
However, the events involved in a Bell inequality test are paired
events detected in coincidence, which means that the subset of the
sample space spanned in the experiment becomes the
intersection\footnote{A pair is counted as such if both Alice AND
Bob register a click simultaneously.} of the two local subsets,
that is $\Omega_{\alpha\beta}=\Omega_\alpha\cap\Omega_\beta$. So,
the combination of two local unfair sampling processes together
with the coincidence condition yields a \emph{global}
contextuality: the sample space $\Omega_{\alpha\beta}$ associated
to the measurement depends explicitly on the entire measurement
context, a feature that is known to compromise the derivation of
Bell inequality \cite{Accardi81,KHR10}. It is interesting to note
that the contextuality of the entire measurement setup arises here
by construction when Bohr had only posited it in his answer
\cite{Bohr} to the challenge to the completeness of quantum
mechanics by Eintein, Podolski and Rosen \cite{EPR35}.

Local realistic models exploiting this detection loophole have
long been proposed \cite{Pearle}, and have been discussed many
times since
\cite{ClauserS,Garra,GargMermin,Santos96,Larsson1,Gisin,Larsson2,PappaR,thompson,AdenierAJP},
but in the absence of any physical justification for the
introduction of such a bias, the existing models using this
loophole are usually deemed to be ad hoc
\cite{Pearle,Larsson1,Gisin}.

We would like to argue that the detectors used in EPR experiments
might very well be responsible for a sample selection bias due to
a threshold effect in their detection capability. Photomultiplier
tubes and avalanche photodiodes, which are commonly used in
quantum optic experiments, are sometimes referred to as
\emph{threshold detectors}\cite{waks} because they cannot
discriminate the number of photons striking a detector
simultaneously. We would like to argue that these detectors can be
called threshold detectors on more account than that. It is known
that there is a fundamental limit on the minimum detectable pulse
energy for a given signal-to-noise ratio\cite{c1,c2,c3}, and
contrary to what is sometimes asserted \cite{BrunnerBG}, we would
like to argue that photomultiplier tubes and avalanche photodiodes
rely in their functioning principle on thresholds as well, and
that it is a feature that should carefully be taken into account
in the framework of quantum information in general, in particular
in the case of a violation of Bell inequalities.

\section{Apparent violation of Bell inequalities with a simple threshold detector}\label{simplemodel}

To illustrate how the existence of a threshold in detectors can be
relevant to the violation of Bell inequalities, let us consider
the simplest possible model of a threshold detector within a
classical framework.

We consider first a source sending pulses impinging on a
polarizing beam splitter (PBS) with its main axis oriented along
the direction $\phi$. Each pulse is assumed to carry the same
energy $E_0$, and the polarization of the pulse is set by a random
variable $\lambda$ uniformly distributed on the interval
$[0,2\pi]$. Two threshold detectors are set at the two outputs of
the PBS (reflected and transmitted). Let $\Phi$ be the detection
threshold. A click is recorded if the portion of the pulse $E$
reaching the detector is greater than this threshold $\Phi$.

By Malus law, the energy reaching the detectors (+) and (-) at the
output of the PBS is:
    \begin{eqnarray}
    E^+&=&E_0 \cos^2(\lambda-\phi)=\frac{1+\cos 2(\lambda-\phi)}{2}
\\
    E^-&=&E_0 \sin^2(\lambda-\phi)=\frac{1-\cos 2(\lambda-\phi)}{2}
    \end{eqnarray}

The measurement results are then:

\begin{itemize}
  \item In the (+) detector channel:
    \begin{itemize}
        \item $A^+=+1$ if $E^+>\Phi$, that is, $\cos
        2(\lambda-\phi)< \frac{2\Phi}{E_0}-1$
        \item $A^+=0$ otherwise
    \end{itemize}
  \item In the (-) channel:
\begin{itemize}
        \item $A^-=+1$ if $E^->\Phi$, that is, $\cos
        2(\lambda-\phi)< -(\frac{2\Phi}{E_0}-1)$
        \item $A^-=0$ otherwise
    \end{itemize}
\end{itemize}

Double clicks occur when both the (+) and (-) channel record a
click simultaneously (for the same pulse), that is, when
\begin{equation}\label{doubleclicks}
    \frac{2\Phi}{E_0}-1<\cos 2(\lambda-\phi)<
-(\frac{2\Phi}{E_0}-1).
\end{equation}

\begin{figure}
\center
\includegraphics[width=12cm]{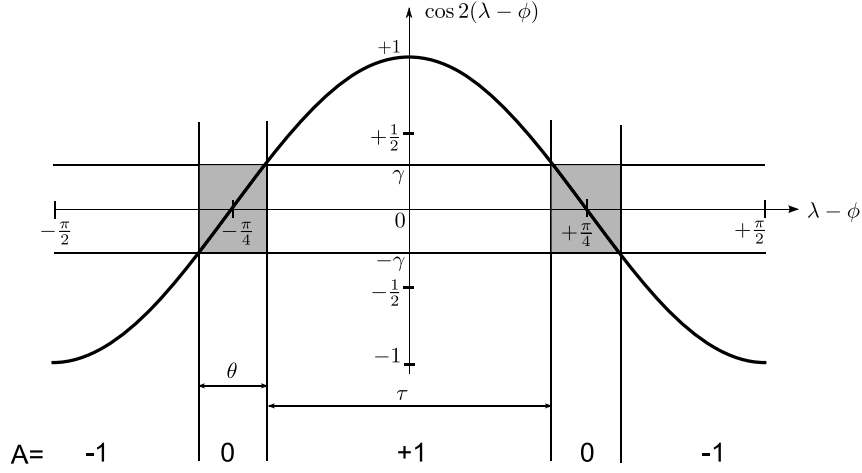}
\caption{\label{detpat}Measurement result as a function of
$(\lambda-\phi)$.}
\end{figure}

We can impose that no double clicks are recorded, by which we mean
that \emph{at most} one of the output channel of the PBS can
record a click for each input pulse. This can be interpreted as
placing ourselves in a `single photon regime' in the framework of
threshold detectors, at least if we stick to the minimalist  maxim
attributed to Zeilinger \cite{QAtoZ}: ``Photons are clicks in
photon detectors.". We will see that this condition is fundamental
to guarantee an apparent violation of Bell inequalities, not only
with this simple model, but with more complex models of threshold
detectors as well. This condition is realized by imposing that the
lower bound in the previous equation is greater than the upper
bound $\frac{2\Phi}{E_0}-1>-(\frac{2\Phi}{E_0}-1)$, or
\begin{equation}\label{nodoubleclicks}
\frac{\Phi}{E_0}>\frac{1}{2}
\end{equation}
Naturally, we also have the condition $\Phi<E_0$ if we want any
click to be recorded at all.

To summarize the behavior of the polarimeter consisting of a PBS
and its two output detectors, we define
\begin{equation}
\gamma=\frac{\Phi}{E_0}-\frac{1}{2},
\end{equation}
with $0<\gamma<\frac{1}{2}$.

The measurement results are displayed in Fig.~\ref{detpat} and can
be written:
\begin{itemize}
    \item $A=+1$ if $\cos 2(\lambda-\phi)>\gamma$
    \item $A=-1$ if $\cos 2(\lambda-\phi)<-\gamma$
    \item $A=0$ if $-\gamma<\cos 2(\lambda-\phi)<\gamma$.
\end{itemize}

Let $\theta$ be the size of the non detection region. We have
$A=0$ when $\arccos(-\gamma)< 2(\lambda-\phi)<\arccos(\gamma$),
that is,
\begin{equation}
\theta=\frac{1}{2}(\arccos(-\gamma)-\arccos(\gamma))=\arcsin(\gamma)
\end{equation}
with the periodicity condition $\tau+\theta=\pi/2$, where $\tau$
is the size each detection region.

We now consider pairs of pulses sent towards two such polarimeters
controlled by Alice and Bob, in a typical EPRB configuration. We
require that the pulses are correlated in polarization, that is
$\lambda_1=\lambda_2=\lambda$, with identical energy $E_0$. We
label $\phi_1$ and $\phi_2$ the measurement settings of the
polarimeters controlled by Alice and Bob respectively, with
$\Delta \phi=\phi_1-\phi_2$.

We can write the correlation as a piecewise function, depending on
the parameter $\Delta \phi$:

\begin{itemize}
  \item $0<|\Delta \phi|\leq\theta$:
\begin{eqnarray}
P^{++}(\Delta \phi)&=&P^{--}(\Delta \phi)=2(\tau-\Delta \phi)\\
P^{+-}(\Delta \phi)&=&P^{-+}(\Delta \phi)=0
\end{eqnarray}
The correlation function is then constant:
\begin{equation}
E= \frac{4(\tau-\Delta \phi)}{4(\tau-\Delta \phi)}=1.
\end{equation}
  \item $\theta<|\Delta \phi|\leq\tau$:
\begin{eqnarray}
P^{++}(\Delta \phi)&=&P^{--}(\Delta \phi)=2(\tau-\Delta \phi)\\
P^{+-}(\Delta \phi)&=&P^{-+}(\Delta \phi)=2(\Delta \phi - \theta)
\end{eqnarray}
The correlation function decreases linearly with $\Delta \phi$:
\begin{equation}
E= \frac{4(\tau-\Delta \phi)-4(\Delta \phi
-\theta)}{4(\tau-\theta)}=\frac{-2\Delta
\phi+\tau+\theta}{\tau-\theta}.
\end{equation}
  \item $\tau<|\Delta \phi|\leq\tau+\theta$:
\begin{eqnarray}
P^{++}(\Delta \phi)&=&P^{--}(\Delta \phi)=0\\
P^{+-}(\Delta \phi)&=&P^{-+}(\Delta \phi)=2(\tau-\theta)+2(\Delta
\phi-\tau)
\end{eqnarray}

The correlation function is then constant:
\begin{equation}
E= - \frac{4(\Delta \phi-\theta)}{4(\Delta \phi-\theta)}=-1.
\end{equation}
\end{itemize}

\begin{figure}
\center
\includegraphics[width=10cm]{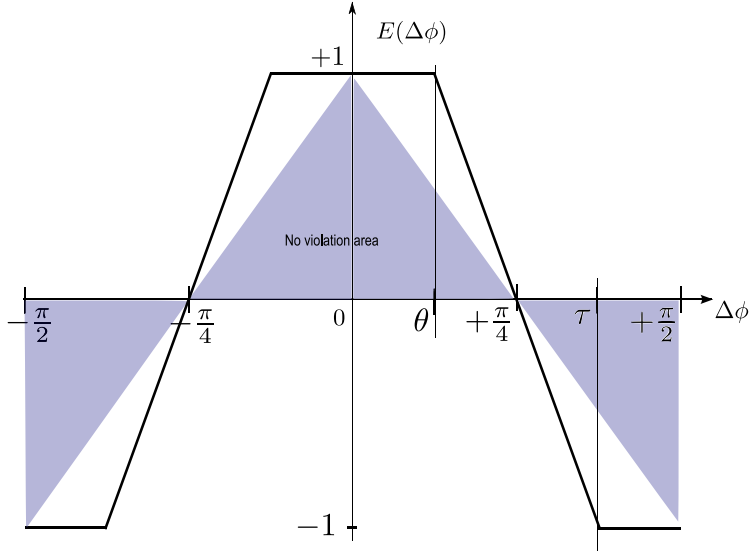}
\caption{\label{corrf1}The correlation function gives an apparent
violation of Bell's inequality. The shaded area represent Bell's
bound: outside this area a violation of Bell inequality is
possible for a well chosen set of measurement settings.}
\end{figure}
The resulting correlation function is displayed in
Fig.~\ref{corrf1}. It is always greater than Bell's bound, and
thus exhibits the appearance of a violates Bell's inequality. The
apparent violation is due to the fact that the probability of
joint detection depends on the angle difference $\Delta \phi$, so
that even though the number of joint detection decreases for small
$\Delta \phi$, the correlation function remains unchanged and of
magnitude 1 for small angle difference (modulo $\pi/2$): it has
the appearance of a stronger correlation than would be classically
possible, but it is of course only apparent and due to an biased
(unfair) sampling process.

For instance, if $\theta>\frac{\pi}{8}$, then $E(|\Delta
\phi|=\frac{\pi}{8})=1$, $E(|\Delta \phi|=\frac{3\pi}{8})=-1$ and
the CHSH function yields:
\begin{eqnarray}
S&=&\big|E(0,\frac{\pi}{8})+E(0,-\frac{\pi}{8})+E(\frac{\pi}{4},\frac{\pi}{8})-E(\frac{\pi}{4},-\frac{\pi}{8})\big|\\
\nonumber &=& 4.
\end{eqnarray}

The correlation function as displayed in Fig.~\ref{corrf1} is
admittedly still far from the result predicted by Quantum
Mechanics, but as we will see below, a slightly more sophisticated
modelization of the threshold detectors can lead to an observed
correlation function in very good agreement with those
predictions.
\section{Photomultipliers and photodiodes as threshold detectors}

%trigger, multiplication, filtering

Photomultipliers and photodiodes come as a large variety of
detectors, with different absorbing materials and designs, but we
would like nevertheless to sketch the characteristics that they
have in common and that are important for our analysis. In
particular, we would like to stress that these detectors rely on
two fundamental thresholds, that we will call the \emph{energy
band threshold} and the \emph{discriminator threshold}.

%%%%%%%%%%%%%%%%%%%%%%%%%%%%%%%%%%%%%%%%%%%%%%%%%%%%%%%%%%%%%%%%%%%%%%%%%%%%%%%%
\subsection{Work function and Energy band threshold}
%%%%%%%%%%%%%%%%%%%%%%%%%%%%%%%%%%%%%%%%%%%%%%%%%%%%%%%%%%%%%%%%%%%%%%%%%%%%%%%

The working principle of photomultipliers and avalanche
photodiodes rely on the excitation of energy carriers bound in a
potential well: in photomultipliers, the energy carriers are
electrons that are extracted from the metal absorber to the
vacuum, whereas in avalanche photodiodes carriers are
electron-hole pairs, the electrons being extracted from the
valence band to the conduction band.

The detection principle of a photomultiplier\footnote{For
simplicity, we will focus here on the case of photomultipliers to
describe the detection process, and we will describe the
differences or similarities in the working principle of
(avalanche) photodiodes when necessary.} is that electrons are
bound in the absorbing material by a potential well, and the
trigger that ultimately leads to a measurable current in the
detector is an electron eventually overcoming this potential
barrier. It is the finite size of the potential well binding the
electron that constitutes this first threshold; the energy
transmitted to the electron has to be high enough to free the
electron from its potential well. In terms of energy band, an
electron susceptible to overcome this potential well is in a low
energy band, where it is susceptible to absorb an incoming
radiation. It has to overcome a band gap in order to reach a
higher energy band (the vacuum) where it can be accelerated enough
to let other electrons overcome the band gap as well. The minimum
required energy is called the \emph{work function} $\Phi$. This
first threshold operates at the input of the detector and can be
referred to as the \emph{band gap threshold}. It implies that such
a threshold detector is practically blind to any incoming
radiation constituted of particles with energy less than that of
the work function. This feature is in fact very useful as it
allows these detectors to be operational for low light signals,
even at room temperature. It prevents the detector from being
blinded by the many fluctuations at lower frequencies than the
targeted signal frequency. This energy threshold was interpreted
by Einstein by quantizing the electromagnetic field and
associating an energy $h\nu$ to a light quanta of frequency $\nu$.
The presence of a minimum energy $\Phi$ allowing for the
photoelectric effect translates into a minimum frequency $\nu_0$.
It should however be remembered that the photoelectric effect can
be explained as a resonance effect by a semiclassical treatment,
in which the field is treated classically while only the energy
levels in matter are quantized \cite{Lamb,LambScully}. We would
therefore like to consider the energy conservation equation used
by Einstein $E=\Phi+E_{\rm kmax}$ independently of his quantizing
hypothesis and consider that $E$ simply represents the energy of a
the electromagnetic pulse striking the detector, and $E_{\rm
kmax}$ the maximum kinetic energy imparted to the liberated
electron (at zero temperature).

The principle of photoemission can be described in three
sequential phases \cite{Photonis,Burles}:

\begin{itemize}
    \item the absorption of a radiation by an electron in the material increases its energy.
    \item the energized electron diffuses through the material, losing
some of its energy by colliding with other electrons (electron
scattering) or with the lattice (phonon scattering), and may reach
the material-vacuum interface.
    \item the electron reaching the surface with enough excess
    energy with respect to the surface barrier that always exist at an interface between material and vacuum may escape from it.
\end{itemize}

Energy losses vary from material to material, but can occur in
each of these steps. In metals, the losses due to electron
scattering are important due to the large number of electrons in
the conduction band, so that in effect the \emph{escape
depth}---the depth at which an electron absorbing a photon may
have a good chance of reaching the surface with enough energy---is
small (only a few nanometers). Electrons absorbing light at
greater depth have practically no chance to reach the surface, so
that increasing the thickness of the photocathode beyond the
escape depth does not result into an increase of efficiency. At
this thickness, photocathode are semi-transparent, so that less
than half the visible light can interact with such a layer. This
puts a practical limit to the quantum efficiency of photocathodes,
independently of the potential barrier.

In semiconducting material, the escape depth is larger due to the
low number of free electrons in the conduction band, which
increases the practical efficiency of such materials.

%%%%%%%%%%%%%%%%%%%%%%%%%%%%%%%%%%%%%%%%%%%%%%%%%%%%%%%%%%%%%%%%%%%%%%%%%%%%%%%%
\subsection{Dark counts and Multiplication noise: necessity of a discrimination threshold}
%%%%%%%%%%%%%%%%%%%%%%%%%%%%%%%%%%%%%%%%%%%%%%%%%%%%%%%%%%%%%%%%%%%%%%%%%%%%%%%

A second threshold, known as the discriminator threshold, arises
due to the inherent noise in threshold detectors, and the
necessity to decide when an output pulse is the signature of a
detected signal, or when it corresponds to noise instead.

The first source of noise is the multiplication process. In
photomultipliers, once an electron has been extracted from the
metal, it is accelerated in vacuum by an electric field until its
kinetic energy is enough to extract other bounds electrons
(secondary emission) when striking the surface of another metal
surface (dynode), which will in turn be used accelerated onto
other dynodes to free more and more electrons, until that flow of
electron ($\sim 10^{7}$) becomes measurable as anodic
current\footnote{ In an avalanche photodiode, a reverse bias is
applied to the P-N junction of the diode so that a high-field is
formed in the depletion layer, where the radiation is absorbed.
Energy carriers (electrons and holes) are accelerated by this
field, and collide with other atoms, liberating secondary carriers
which are in turn accelerated and collide with other atoms
(avalanche effect).}.

The multiplication factor is not a constant, it varies
significantly from one liberated energy carrier to another. In a
photomultiplier, the gain is given by
\begin{equation}
g=\alpha \delta^N
\end{equation}
where $\delta$ is the multiplication factor of a single dynode,
$\alpha$ is the fraction of photoelectrons collected by the
multiplier structure, and $N$ is the number of stages in the
photomultiplier. In the most simple model, $\delta$ can be assumed
to follow a Poisson distribution about the average yield for each
dynode, so that the gain is a compound Poisson process over $N$
identical stages. However, experimental measurements of the single
photoelectron pulse height spectra from photomultipliers exhibit a
distribution with larger relative variance than predicted by the
Poisson model (and in some case with an decreasing exponential
distribution instead of a peaked one), and there is thus no
universal description of multiplication statistics \cite{Knoll}.

The second source of noise is due to the fact that even in total
darkness a current can still be measured at the output of a
photomultiplier or of a photodiode. This background is referred to
as the \emph{dark current}, or \emph{dark counts} (depending on
the operating mode of the detector). It is due to various sources
of identified noise, such as ohmic leakage, regenerative effects,
and cosmic radiations; but the main source of noise under normal
condition is thermionic emission.

As we have seen, a ionization occurs when the energy of the
electron is high enough to let it overcome this potential barrier.
This can naturally happen when an incoming radiation excites an
energy carrier, but this can also happen through thermal
fluctuation even in the absence of any incoming signal. Even
though the probability of having an electron spontaneously
overcoming the band gap due to thermionic energy is very small (as
$kT\ll \Phi$ at room temperature), the large number of free
electrons in the absorber means that this will in effect lead to a
significant number of thermionic pulses in the output of the
detector. The amplitude distribution of the dark current varies
according to the type of dynode, and may vary considerably between
different samples of the same photomultiplier. However, because a
thermionic photoelectron can appear at any stage in the
multiplication process, the background spectrum typically has a
high proportion of undersized pulses (fractional photoelectrons).

The combination of thermionic emissions with the stochastic nature
of the multiplication leads to output pulses that have a large
dispersion of shape and amplitude, with an excess of small
amplitude pulses.

Ideally, one would wish to be able to subtract these unwanted
small pulses from the signal that is to be measured by subtracting
this background from the measured signal plus background. In low
light applications, the optimum sensitivity is obtained in photon
counting mode. The idea is to eliminate the effect of varying
pulse heights by using a \emph{discriminator}. A discriminator is
a circuit that produces a specified output signal if and only if
it receives an input pulse whose amplitude exceeds in one case or
is less than an assigned value in another case\cite{Burles}. It
produces a standard pulse for all photomultiplier pulses above a
fixed threshold \cite{Wright1}. It is only when the output current
is greater than this low level discrimination value that a pulse
is \emph{discretized} as a valid ``click".

Setting the value of this discriminator threshold is a delicate
point. A too low threshold will pick up many dark counts, thus
degrading the signal and the correlation in coincidence counting.
A higher level threshold will improve the signal-to-noise ratio,
but at the cost of rejecting many low pulses, thus decreasing the
efficiency of the detector \cite{AspectTh,GrangierTh,Wright1}.
Knowing how the efficiency of detectors is a crucial weakness in
EPR-Bell experiments, the influence of this discriminator
threshold should be considered with the utmost attention. When the
output distribution of single photoelectron has a peak
distribution, one would ideally set the threshold in the valley
right before the peak, but this requires to plot a pulse height
distribution and is not always practical. One possibility consists
in plotting the count rate as a function of the threshold for a
fixed high voltage, and chose a value on the observed plateau
\cite{AspectTh}. There is however a level of arbitrariness in this
choice. For instance, Aspect \cite{AspectTh} chose a value in the
center of the plateau, with a higher discrimination that was
possible, but with the advantage of stability (in the presence of
variations in the multiplication voltage), whereas Grangier
\cite{GrangierTh} chose a higher threshold, at the cost of lower
efficiency, but with less spurious counts.

In the literature on the experimental violation of Bell
inequalities, the choice of the discriminator threshold is at best
discussed after the theoretical discussion, as a merely technical
detail that only influences the quantum efficiency of the detector
thought as black box.  We believe that the discrimination
threshold should be included in any discussion involving detection
of weak light pulses because it can have a manifest effect on
statistics that are investigated. Indeed rejecting low pulses may
very well introduce a sample selection bias, favoring the
appearance of non classical effects such as an apparent violation
of Bell inequalities.
%%%%%%%%%%%%%%%%%%%%%%%%%%%%%%%%%%%%%%%%%%%%%%%%%%%%%%%%%%%%%%%%%%%%%%%%%%%%%%%%
\section{Simulations with classically correlated pairs of pulses}\label{simulation}
%%%%%%%%%%%%%%%%%%%%%%%%%%%%%%%%%%%%%%%%%%%%%%%%%%%%%%%%%%%%%%%%%%%%%%%%%%%%%%%
To sum up, the two thresholds that we wish to consider in
photomultipliers and photodiodes in the context of EPR-Bell
experiments are:
\begin{itemize}
  \item Energy gap threshold (by which the low energy \emph{inputs} pulses are discarded)
  \item Discriminator threshold (by which the low amplitude \emph{outputs} pulses are discarded)
\end{itemize}

We believe that the influence of these thresholds may been
overlooked in the discussion on non-classical effects, and in
particular in EPR and Bell inequality discussion. Individually,
each threshold seems quite harmless, but we will show here that
their \emph{combined} effect can have a significant influence on
the coincidence counting statistics. Indeed the combined effect of
these two threshold opens the possibility of a fundamental, and
yet tunable, detection threshold in photon counting, which can
lead to a sample selection bias in the detected events, leading to
an apparent violation of Bell inequalities in a fully local
realistic framework.

We have performed numerical simulations in which a pair of pulses
with same energy and polarization are sent over arbitrary
distances. Alice and Bob make a polarization measurement using
polarizing beam splitter (PBS). Two threshold detectors record the
`clicks' at the output of the two possible output (reflected and
transmitted) of each PBS. They count the number of coincidences
and compute the correlation function.

A source creates pairs of classical laser pulses with identical
linear polarization $\lambda$:
\begin{itemize}
    \item The polarization of pairs is uniformly distributed over $[0,2\pi]$.
    \item The polarization measurements are performed with polarizing beam spitters $\varphi_1$ and $\varphi_2$.
    \item The collection efficiency is set to 1 for simplicity. Each pulse is fully absorbed by an energy carrier.
    \item The detectors in each of the four channels are modelized with varying
    \begin{itemize}
      \item Ratio of pulse energy over the band gap threshold
      \item Discrimination threshold
    \end{itemize}
\end{itemize}

In the simplest version of our simulation, we have used some
simplifying assumptions that are quite unrealistic, but that
nevertheless show the essential feature of the problem. We have
assumed that each pulse carries the same energy $E_0$, that the
pulses are well separated in time so that only one pulse reaches a
detector at a time, and that the energy $E$ reaching a detector is
fully absorbed by one single energy carrier. We have also assumed
that the potential well binding the energy carrier has a well
defined and fixed size $\Phi$, so that the energy conservation is
written $E=\Phi+E_k$, where  $E_k$ is the excess energy of the
carrier after absorption. The multiplication was flawless (the
gain $g$ was set to 1), so that the output current $i=g.E_k$  and
the detector was exempt of dark counts. The discriminator $D$ is a
logical gate evaluating the proposition $i>D$. In the absence of
noise and with a gain of 1, increasing the discrimination
threshold $D$ is simply equivalent to increasing the potential
well binding the energy carriers by an amount $D$.

\begin{figure}
\center
a)
\includegraphics[width=7.5cm]{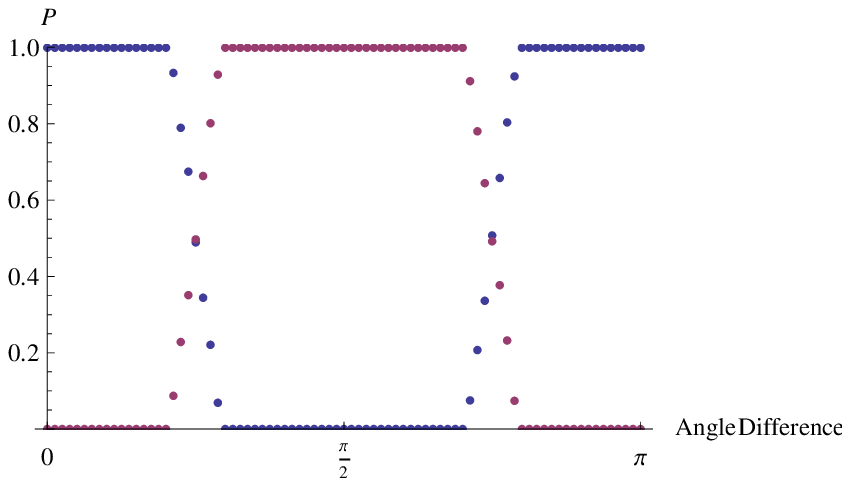}
b)
\includegraphics[width=7.5cm]{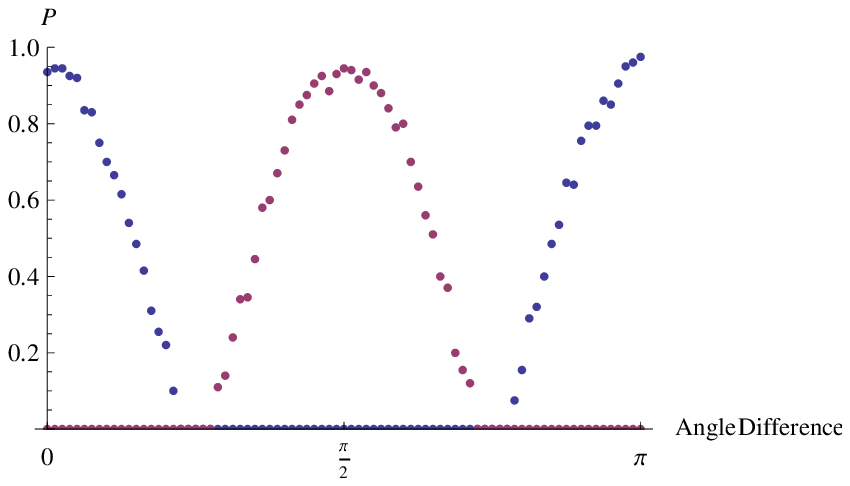}
\caption{\label{fig:detpatAB}An actual detection probability as a
function of the difference ($\lambda-\varphi$). a) for Alice with
$\frac{E_0}{\Phi}=2.3$, b) for Bob with $\frac{E_0}{\Phi}=1.9$. In
both cases, the discriminator threshold was set to
$D_{\textrm{threshold}}=0.3$ (with collection efficiency and gain
set to 1). The pattern is similar to the Gisin-Larsson model
\cite{Gisin,Larsson1,Larsson2}.}
\end{figure}

\begin{figure}
\center
a)
\includegraphics[width=7.5cm]{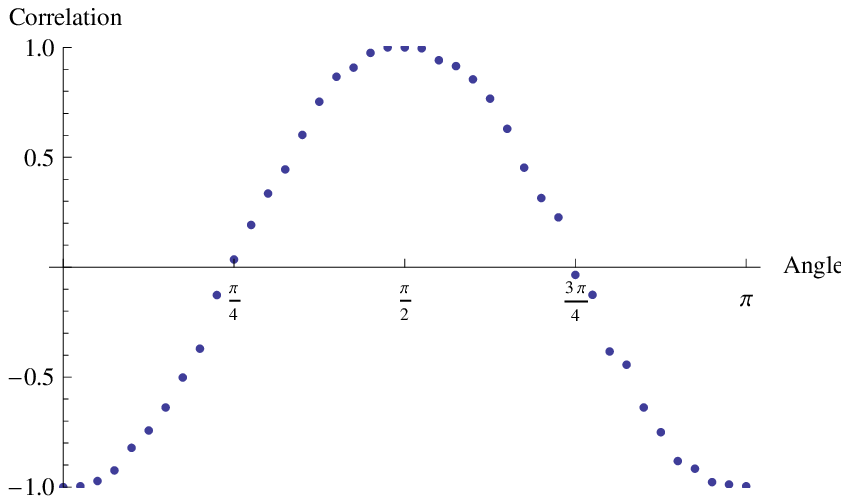}
b)
\includegraphics[width=7.5cm]{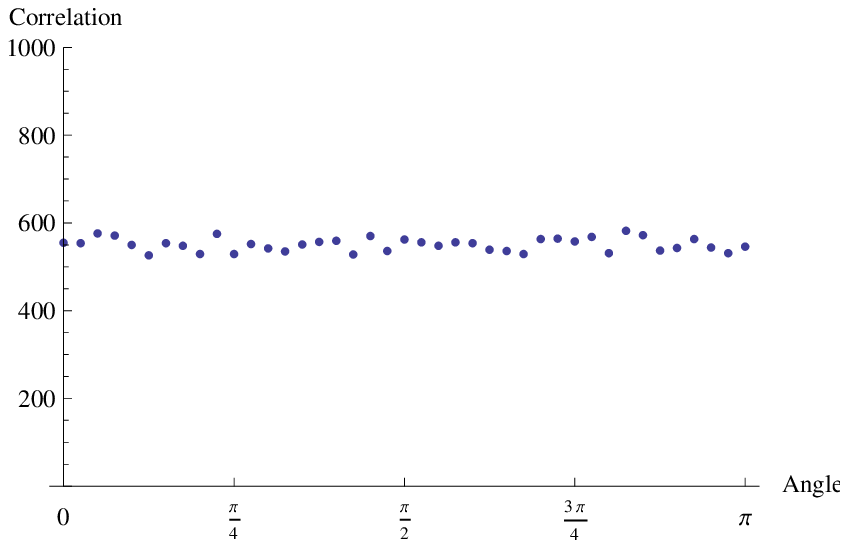}
\caption{\label{fig:correlationAB}Measured results with the
settings of Fig.~\ref{fig:detpatAB}. a) The correlation function
is in excellent agreement with the $\cos(2\Delta \phi)$ predicted
by Quantum Mechanics. b) Total number of coincidences}
\end{figure}

In order to get closer to the features of real detectors, we have
considered the fact that the work function is a \emph{minimum} for
the potential well seen by the energy carrier, which can be
modelized by introducing some losses in the absorption of the
incoming energy by the carrier. The losses alone in the absorbed
energy are enough to make the model flexible enough to recover the
predictions of Quantum mechanics, by using non symmetrical
detection pattern. This is done settings having
 a different ratio between the energy of the incoming pulse $E_0$ and
 the work function $\Phi$ at the two stations controlled by Alice and
 Bob, which effectively changes the efficiency at each station. One can
 then obtain detection patterns pretty similar to that of the Gisin-Larsson
  model \cite{Gisin,Larsson1,Larsson2} (see Fig.~\ref{fig:detpatAB}),
  leading to a correlation function and a total number of counts that mimics
   the predictions of Quantum mechanics rather nicely (see Fig~.~\ref{fig:correlationAB}).

\begin{figure}
\center
a)
\includegraphics[width=7.5cm]{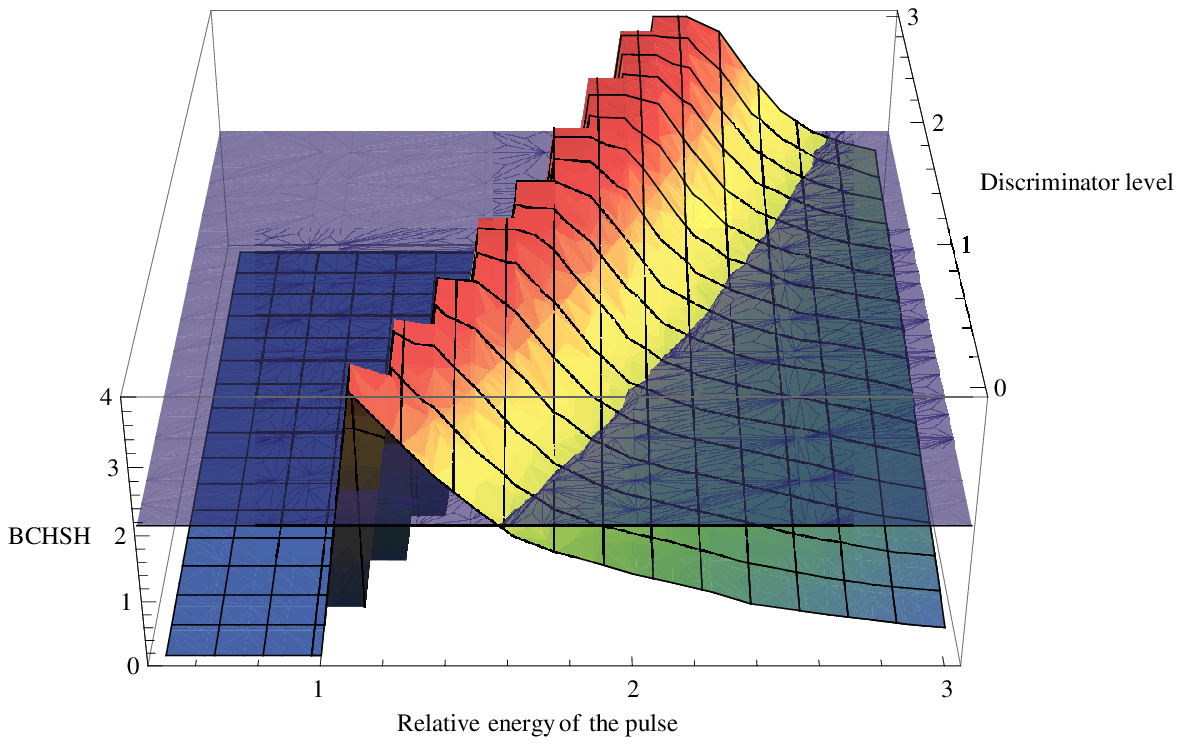}
b)
\includegraphics[width=7.5cm]{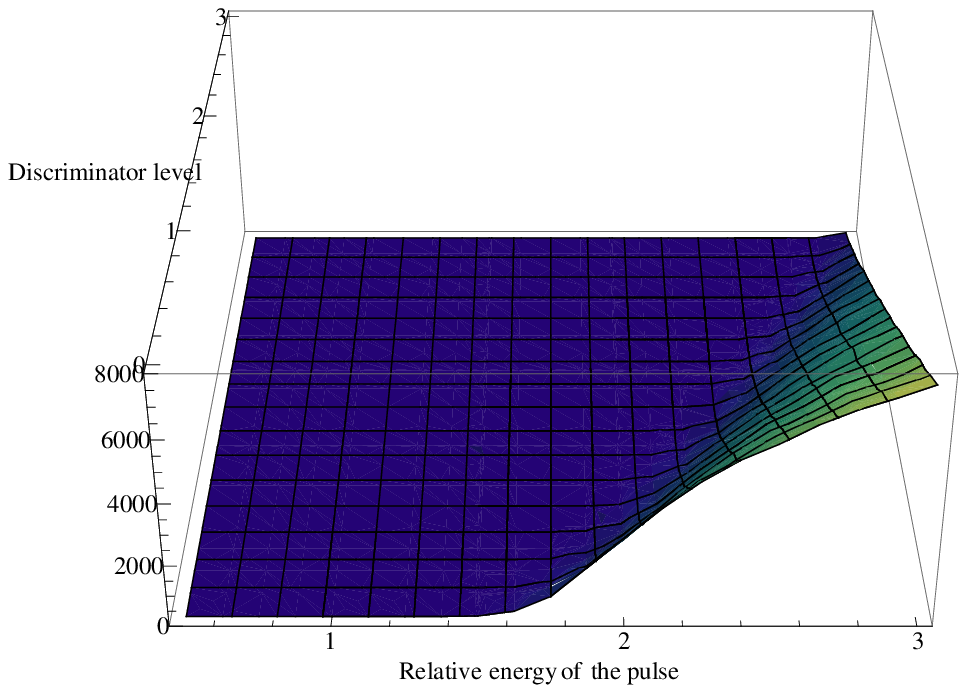}
\caption{\label{fig:islandBCHSH}a) (Apparent) violation of Bell
inequalities as a function of the relative energy of the pulse
$E_0/\Phi$ and of the discrimination threshold $D$ (horizontal
coordinates). The flat semi transparent horizontal layer at
BCHSH=2 represents Bell inequalities. An apparent violation is
observed for a combination of the two threshold.b) Corresponding
Number of double counts. Having coincidence clicks between the $+$
and $-$ channel at each station should be avoided to remain in the
`single photon regime'. One is thus necessarily driven to regions
where an apparent violation of Bell inequalities is observed.}
\end{figure}

By varying simultaneously the ratio $E_0/\Phi$, where $E_0$ is the
energy of the incoming pulse $E_0$ and $\Phi$ the work function,
together with the discrimination threshold $D$, we can obtain a 3D
map showing for which couple of parameters an apparent violation
of Bell inequalities is obtained (see Fig.~\ref{fig:islandBCHSH}),
as well as the corresponding total number of coincidences. The
result shows a significant region where an (apparent) violation of
Bell inequalities is obtained. The magnitude of the violation
increases as $E_0/\Phi$ decreases and as $D$ increases. It reaches
the maximum 4 just like our simple model of Section
\ref{simplemodel} did, which is hardly surprising given the
simplicity of the model.

Note that the number of double counts, that is, the number of
coincidence counts between two detectors at the same station
(either Alice or Bob), is a serious constraint on the combination
of parameters $E_0/\Phi$ and $D$ that can be considered
acceptable. Operating away from the region where an apparent
violation of Bell inequalities arises from the combined threshold
cannot be done without having double counts appearing (see
Fig.~\ref{fig:islandBCHSH}b). Just like in our simple model of
Section \ref{simplemodel}, the constraint of having no double
counts necessarily leads to an apparent violation of Bell
inequalities.

We have performed more complex simulations with the implementation
of various sources of noise, losses and fluctuations
\cite{AdenierPhD}. The fundamental result was unchanged. Naturally
the noisier the simulation, the lesser the apparent violation of
Bell inequalities. Introducing some multiplication noise and some
dark counts diminishes the maximum observable, or at the very
least it is only with a rather high discriminator threshold $D$
that such high values can be obtained. It also naturally increases
the number of small output pulses, and thus necessitates to keep
the discriminator threshold above a certain level to suppress the
double counts, a level that is always concomitant with an apparent
violation of Bell inequalities.

\section{Discussion}

The result of the simulation depends crucially on the classical
behavior of a pulse at a polarizing beam splitter. We have indeed
assumed throughout this article that pulses can be treated
classically. At a PBS oriented along $\phi$, the energy of the
pulse linearly polarized along $\lambda$ is split in both arms
depending on $(\lambda - \phi)$, according to Malus law. It should
be noted that the assumption that the field would be classically
split at a PBS need not be in contradiction with known
experimental facts, since many phenomena can be explain without a
field quantization. For instance, it it well known that the
photoelectric effect can be explained without resorting to the
concept of photon\cite{Lamb}, by considering classical fields and
quantized energy levels in atoms \cite{LambScully}. In fact, it
looks like that the only type of effects that are considered to
require such a field quantization are of a similar type as the
EPR-Bell experiments, which, as we have seen, can actually be
explained with threshold detectors and classical fields.

The other possibility is of course to consider that each pulse
contains one indivisible quantum of light (photon), each of them
being either fully reflected or fully transmitter at a PBS, the
which-way information being determined by Malus law, this time
understood as a probability rule. In such a case, regardless of
$\lambda$ and $\phi$, the detector always sees the same energy in
the single photon regime, because a photon is either completely
transmitted or completely
reflected\cite{Dirac,BrukZeil1,BrukZeil2,Whooters}, so that no
sample selection bias can arise. It means that the role of the
thresholds as a possible source of unfair sampling can only be
neglected in an EPR-Bell setup if one assumes that the light
impinging on a polarizing beam splitter is constituted of
indivisible particles (light quanta). The assumption of fair
sampling would thus be an assumption on the indivisibility of
light quanta in disguise. This is usually considered the only
possible interpretation of the anticorrelation experiments at a
beam splitter \cite{Grangier86,Jacques}, but we would like to
stress that this result can be enforced and reproduced
classically, as soon as a detection threshold are considered, as
was is shown above with Eqs. (\ref{doubleclicks}) and
(\ref{nodoubleclicks}).

\section{Conclusion}

We have seen that the role of thresholds in photomultipliers and
avalanche photodiodes is fundamental for the interpretation of
EPR-Bell experiments, but this should also be true for other
non-classical experiments in which the only difference between the
classical case and the quantum case is the visibility of the
interference effect, as was shown to be the case for all type of
two-photon experiments by Klyshko \cite{Klyshko,Klyshko2}. Indeed,
as we have shown here in the case of EPR-Bell experiments, the
introduction of threshold detectors in a classical framework has
the simple effect of lowering the visibility of the interferences
that would be obtained without such thresholds, that is, with
detectors that would offer a linear response regardless of the
incoming signal. Fair sampling should therefore no longer remain
as an unproven assumption in nonclassical experiments, it is high
time for this fundamental loophole to be thoroughly tested
experimentally \cite{AdenierJRLR} with the same attention and care
that has been devoted to the communication loopholes.

%\begin{thebibliography}{9}
\bibliographystyle{unsrt}
\bibliography{refs2}

% Stop your text
\end{document}